\documentclass[12pt]{iopart}
\usepackage{epsfig}
\usepackage{graphicx}
\usepackage{mathabx}
\usepackage{lscape}
\usepackage{longtable}

\begin{document}

\title[A 40 MYR OLD GASEOUS DISK]{A 40 MYR OLD GASEOUS CIRCUMSTELLAR DISK AT 49 CETI:
MASSIVE CO-RICH COMET CLOUDS AT YOUNG A-TYPE STARS}

\author{B. Zuckerman$^1$ and Inseok Song$^2$}

\address{$^1$Department of Physics and Astronomy, University of California, Los Angeles, CA 90095, USA}
\address{$^2$Department of Physics and Astronomy, University of Georgia, Athens, GA 30602-2451, USA} 
\eads{\mailto{ben@astro.ucla.edu,song@physast.uga.edu}}
\begin{abstract}

The gaseous molecular disk that orbits the main sequence A-type star 49 Ceti has been known since 1995, but the stellar age and the origin of the observed carbon monoxide molecules have been unknown.  We now identify 49 Ceti as a member of the 40 Myr old Argus Association and present a colliding comet model to explain the high CO concentrations seen at 49 Ceti and the 30 Myr old A-type star HD 21997.  The model suggests that massive -- 400 Earth mass -- analogs of the Sun's Kuiper Belt are in orbit about some A-type stars, that these large masses are composed primarily of comet-like objects, and that these objects are rich in CO and perhaps also CO$_2$.  We identify additional early-type members of the Argus Association and the Tucana/Horologium and Columba Associations; some of these stars display excess mid-infrared emission as measured with the Widefield Infrared Survey Explorer (WISE).

\end{abstract}
\maketitle

\section{INTRODUCTION}

Extensive observations have shown that by the time a typical T Tauri star is $\sim$5 Myr old its protostellar disk will retain insufficient carbon monoxide gas to be detectable with a radio telescope.   Stars with ages $<$5 Myr are virtually non-existent within $\sim$100 pc of Earth (Zuckerman \& Song 2004; Torres et al 2008).   Thus, stars this close to Earth with detectable gas-rich circumstellar disks are few and far between.  CO has been detected from three classical T Tauri stars within 100 pc of Earth:  TW Hya, V4046 Sgr and MP Mus (Kastner et al 2010; Rodriguez et al 2010; and references therein).   Such stars, with likely ages in the range 7-12 Myr, represent the oldest known examples of remnant gaseous protoplanetary disks around solar-type stars.

Two dusty A-type stars near Earth with gas-rich disks -- 49 Cet and HD 21997 -- are quite different from classical T Tauri stars.  The circumstellar gas at 49 Cet (= HR 451 \& HIP 7345) was discovered in 1995 and has warranted the attention of single and interferometric radio telescopes (Zuckerman et al 1995; Dent et al 2005; Hughes et al 2008).  The dust distribution has been investigated with high spatial resolution mid-infrared imaging (Wahhaj et al 2007).  The gas around HD 21997 (= HR 1082 \& HIP 16449) was discovered much more recently (Moor et al 2011).  49 Cet and HD 21997 are, respectively, 59 and 72 pc from Earth (van Leeuwen 2007). 

HD 21997 is classified, reasonably securely, as a member the 30 Myr old Columba Association (Moor et al 2006; Torres et al 2008).   The age of 49 Cet has been much more up in the air.  The earliest estimates of this age were typically 8-10 Myr and 8 Myr is the age adopted in the interferometric study by Hughes et al (2008).  With such a young age, no older than that of the three classical T Tauri stars mentioned above, it was reasonable to interpret 49 Cet as a star in transition between a protoplanetary and a debris disk.   However, largely because they did not associate the Galactic space motions (UVW) of 49 Cet with any known young (8-100 Myr old) stellar moving group, Rhee et al (2007) tentatively assigned an age of "20?" Myr to the star.

Based on the discussion that follows below, we believe that all previous age estimates for 49 Cet are too young and that the star is actually a member of the 40 Myr old Argus Association (Torres et al 2008; Zuckerman et al 2011).   Thus among main sequence stars, 49 Cet and HD 21997 contain the longest-lived substantial reservoirs of circumstellar gas currently known in astronomy.  Moor et al (2011) outline how difficult it is to explain how so much molecular gas can be present in the circumstellar disks of such old stars.  In the present paper we present
a model that appears capable of accounting for the properties of such stars.

\section{THE AGE OF 49 CET}

We deduce the age of 49 Cet based on its Galactic space motions (UVW) and its location on an A-type star color-magnitude diagram (CMD).  Together these demonstrate that 49 Cet is a member of the Argus Association whose age, based on analysis of its G- and K-type members (Torres et al 2008), is $\sim$40 Myr.

Table 1 lists A-type stars that we propose as likely or possible members of the Argus Association; some of these stars are considered in Section 3.   The mean UVW of the Association is given in the Table notes.  As may be seen, the measurement errors in radial velocity for some Table 1 stars are substantial; until more accurate velocities are measured, identification of such stars with Argus can be regarded only as possible.  However, the listed heliocentric radial velocity (10.3$\pm$0.7 km s$^{-1}$) for 49 Cet (= HIP 7345) is well determined (Gontcharov 2006) and agrees well with the systemic velocity (12.2 km/s) determined from the J = 2-1 line of CO (Hughes et al 2008).   While measurement of accurate radial velocities for rapidly rotating early-type stars is challenging, precise velocities can be easily measured for radio emission lines from circumstellar disks.  Thus the Galactic space motion of 49 Cet is consistent with that of the Argus association.   From a list of $\sim$1500 early-type main sequence stars within 80 pc of Earth very few are found to have all three components of UVW close to those of the Argus Association (six such stars are listed in Table 4 of Zuckerman et al 2011).

A star is included in the present Tables 1 and 2 or in Tables 4-6 of Zuckerman et al (2011) if each component of UVW is consistent with that of the association mean.  "Consistency" in this context implies that all 3 stellar components of UVW match those of the association mean within the sum of the quoted error in stellar velocity and the standard deviation of the dispersion among the association velocities.  However, because errors in UVW can be underestimated (see following paragraph), because a standard deviation is not an absolute limit, and because these Tables list "proposed" association members, we have tended to err somewhat on the side of including a potential non-member in preference to losing a true member.  

In Figure 1 we plot the location on a CMD of early A-type stars near Earth that have been proposed as members of five young moving groups.  Table 2 presents early-type stars that we propose as members of the 30 Myr old Tucana/Horologium and Columba Associations; membership is based on UVW and location on a CMD.
As may be seen from Figure 1 and Table 3, for a given B-V there sometimes is a substantial spread in M$_V$ for stars proposed to be of the same or nearly the same age.  If the stars have been correctly placed into the indicated moving groups, then one possibility is that some aspect of the formation and early evolution of early-type stars can lead to a range of stellar sizes at a given color and nominal age.  Alternatively, some of the stars in Table 3 may not actually belong to the indicated moving group.  Given the errors in measuring radial velocities of early A-type stars and, to a lesser degree, errors in Hipparcos parallaxes, some proposed members of a given moving group will ultimately turn out to be non-members.  In the present work and that of our 2011 paper (Zuckerman et al 2011) we have adopted the Gontcharov (2006) radial velocities and their errors while recognizing that the latter are likely to be too optimistic for some stars.   We adopt Hipparcos parallaxes from van Leeuwen (2007) while anticipating that some of these may not be correct (see, for example, the example of HIP 30675 considered in Section 3).  Ultimately, high accuracy parallaxes from Gaia, a dedicated campaign to confirm the Gontcharov radial velocities and reduce their errors, and perhaps a traceback analysis of the moving groups, for example, as demonstrated by Ortega et al (2002) for the $\beta$ Pictoris moving group, should distinguish between true members and impostors.

Notwithstanding caveats expressed in the preceding paragraph, the excellent agreement of the UVW of 49 Cet with that of the Argus Association and the location of 49 Cet (the red square in Figure 1 and star A in Table 3), indicates that 49 Cet is unlikely to be younger than 40 Myr.  Indeed from comparison on Figure 1 of the location of the various proposed A-type members of Argus with the indicated Pleiades stars (from Vigan et al 2012), the Argus Association might be somewhat older than the 40 Myr deduced by Torres et al (2008). 
Thus, a 40 Myr age for 49 Cet is at least as secure as the 30 Myr age for HD 21997 deduced by Moor et al (2011).   HD 21997 is star "E" in Figure 1.

\section{YOUNG NEARBY STARS OTHER THAN 49 CET}

Various searches for CO toward dusty main sequence stars with ages $>$10 Myr have been reported, but only 49 Cet and HD 21997 have yielded detections (e.g., Zuckerman et al 1995; Dent et al 2005; Kastner et al 2010; Moor et al 2011).  In Appendix A and Table 4 we present a revised version of a portion of Table 1 in Kastner et al (2010).  A principal motivation for revisiting this table is the absence of CO abundance upper limits in the Kastner et al table which presents only limits to H$_2$ abundances based on an assumed H$_2$/CO ratio by number equal to 10$^4$.  As discussed in Section 4.3 below, if the comet model described in Section 4.4 can account for the CO observed at 49 Cet and HD 21997, then the H$_2$/CO ratio in these debris disks is unconstrained and is unlikely to be 10$^4$.

In the present section we consider mid IR emission as measured with the Widefield Infrared Survey Explorer (WISE), UVW velocities, and location on a color-magnitude diagram of some stars listed in Tables 1-3 along with a few additional stars of interest.

\subsection{Stars With Emission In the WISE Catalog}

We examined the WISE catalog to determine if any stars in Tables 1 and 2 had excess IR emission above the stellar photosphere.  Our criteria for deciding whether an excess is present are the same as described in Section 2.2 of Zuckerman et al (2012), so we do not repeat these criteria here.
Four stars from Tables 1 and 2 display excess emission above the stellar photosphere in the WISE 22 $\mu$m filter; their spectral energy distributions are shown in Figure 2.

\subsection{UVW and Location On Color-Magnitude Diagrams}

HIP 23585: This Table 1 star is too red to appear in Figure 1, but can be plotted on similar color-magnitude diagrams that appear in Zuckerman (2001) and Vigan et al (2012).  Its location is consistent with a 40 Myr age.

HIP 25280, 59394, 72408, 104365:  Based on their W component of space velocity, these Table 2 stars are members of the Tuc/Hor Association (rather than Columba).

HIP 26309: This member of the nucleus of the Columba Association (Zuckerman et al 2011), sits very low on a CMD (star H in Figure 1).

HIP 26395 and 53771: Based on their W component of space velocity, these Table 2 stars are members of the Columba Association (rather than the Tuc/Hor Association).  HIP 26395 is a new member of the nucleus of the Columba Association; the nuclear region is delineated in Section 5.2 of Zuckerman et al (2011).
HIP 26395 is star "I" in Table 3 and Figure 1; it, along with another Columba member, HIP 32104 (star K), and a Tuc/Hor member HIP 12413 (star C) all sit near each other and very low on the CMD.   

HIP 30675:  This $\sim$20 arcsec double star, HR 2328, is potentially nearby and young.  According to SIMBAD and van Leeuwen (2007), the parallax is 13.26 mas which would put the star $\sim$75 pc from Earth.  However, there are other sources where the listed parallax is much smaller, near 8 mas; for example in the Hipparcos double star catalog and in Kharchenko \& Roeser (2009).  If 13.26 is to be believed, then the star plots too low in Figure 1 to be reasonable, whereas if something like 8 is used, then the location is reasonable.   Also, the small proper motion is more likely for a more distant star.   At a distance of 125 pc, the UVW is -10.0,-19.4,-5.7 (1.9,2.0,1.3), consistent with membership in the Columba Association (see Notes to Table 2).  The primary is A0-type and the secondary about F5.  Even for a young star, the F-star is very bright in X-rays according to Figure 4 in Zuckerman \& Song (2004); based on the ROSAT All-Sky Survey, its fractional X-ray luminosity (L$_x$/L$_{bol}$) is equal to -3.8.

HIP 41081: The U velocity component is that of Tuc/Hor while the W component is appropriate for Columba.  Similar mixed-messages occur for a few stars in Table 3 of Zuckerman et al (2011).   

HIP 53771: While too red to appear in Figure 1, the location of this Table 2 star on a CMD  (Zuckerman 2001; Vigan et al 2012) is plausible for a 30 Myr old star.

HIP 89925:  This double line spectroscopic binary consists of Am and A9 stars (Fekel et al 2009).  A deconvolution of the light curve, combined with assumed B-V = 0.14 and 0.26 for the primary and secondary respectively (Fekel et al 2009), places both components in the region of a CMD (Zuckerman 2001; Vigan et al 2012) appropriate for 40 Myr old stars.

\section{DISCUSSION} 

What is the origin of the molecular gas at 49 Cet and HD 21997?  Some 
possibilities include (1) unusually long-lived remnants of the 
protoplanetary (Herbig Ae) phase of stellar evolution, (2) 
collisions of icy comets, and (3) 
signature of the aftermath of a recent collision of gas-rich planets or 
planetary embryos.

The lifetime of a given CO molecule in a circumstellar disk is set by
the intensity of the ultraviolet stellar and interstellar radiation
fields modulated by shielding by other CO molecules and by H$_2$ molecules
and dust grains.  In their model for HD 21997, Moor et al (2011) derive
a CO lifetime of $\sim$500 years and, given similarities in the two debris
disks, the CO lifetime at 49 Cet is unlikely to be much different.  500 years is
so short compared with estimated ages of 30-40 Myr that we concur with Moor et al
that the observed CO is  not likely to be a remnant of the protoplanetary
phase.  Thus, in what follows, we consider models with rapid production
of gas phase CO.  We assume planetesimal belts that contain
volatile-rich comet-like and larger objects, akin to the Sun's Kuiper Belt.

Moor et al (2011) and Hughes et al (2008) find that shielding of CO by 
dust grains is unimportant in comparison with shielding by H$_2$ or CO 
self-shielding.  Nonetheless, the dust abundance in the debris disk can 
potentially yield important clues regarding the origin of the CO 
molecules.  For example, 49 Cet and HD 21997 are among the dustiest A-type 
stars within 70 pc of Earth thus suggesting a direct connection between dust and gas 
even if the former does not protect the latter.

To determine how the amount of dust at 49 Cet and HD 21997 compares with that at other 
A-type stars near Earth, we consider stars with excess 60 $\mu$m measured with the Infrared 
Astronomical Satellite (IRAS) and listed in Table 2 of Rhee et al (2007).  This
table should be complete for A-stars within 70 pc of Earth with large fractional infrared luminosities, $\tau$ = 
L$_{IR}$/L$_{bol}$;
all such stars would have been easily detectable with IRAS.  According to Rhee et al (2007), $\tau$ = 8 x 10$^{-4}$
and 5 x 10$^{-4}$ for 49 Cet and HD 21997, respectively.  There are $\sim$300 A-type stars
within $\sim$70 pc of Earth (Jura et al 1998).  Of these only 4 have $\tau$ as large or larger than those of 49 Cet and HD 21997.
Three of the four -- HR 4796, $\beta$ Pictoris, HR 7012 -- have ages $\sim$10 Myr and may not yet have
completely dispersed their primordial dusty disks.   In addition, the dust at these three stars is warmer than
the dust at 49 Cet and HD 21997, and the closer to a star that dust grains reside the fewer are required to absorb a 
given fraction of the starlight.   The remaining very dusty star is HIP 8122 with $\tau$ = 5 x 10$^{-4}$ and an estimated age of
100 Myr (Rhee et al 2007).  Thus, for stars whose infrared emission is dominated by Kuiper Belt analog, cool dust, 
49 Cet and HD 21997 rank among the 1-2\% dustiest of nearby A-type stars.  (We do not consider here the tiny number
of young nearby A-type stars with IR-excess emission dominated by warm, zodiacal-type dust (Melis et al 2012)).

Youth and dustiness notwithstanding, CO has not yet 
been detected with a radio telescope at any of HR 4796, $\beta$ Pictoris, or HR 7012, consistent with the 
notion that the CO gas seen at 49 Cet and HD 21997 is not primordial.  If the CO gas were primordial, then on average one would expect more CO in a dusty 10 Myr old 
circumstellar envelope than one of age 30-40 Myr.  While there have been detections via optical absorption lines of trace amounts of gas around some main sequence stars including $\beta$ Pictoris (e.g., Roberge \& Weinberger 2008 and references therein), the derived column densities of gas are far less than those necessary for radio detection of a rotational line in emission.

The far infrared emission seen by IRAS is carried primarily by grains of radius $\sim$10 $\mu$m, just above the radiative blowout size.  Flux densities 
at submillimeter wavelengths sample larger grains, with radii up to a few mm.   The total mass of these large grains has been 
measured with radio telescopes at various stars; results are listed in Table 2 in Rhee et al (2007) and -- for 
Vega, Fomalhaut, 49 Cet, and HD 21997 -- are
presented in the 4th column of Table 5 of the present paper.  Again, whether one 
examines Table 2 in Rhee et al (2007) or Table 5 here, 49 Cet and HD 21997 stand out for having large total dust masses in comparison 
with other nearby stars. 

If CO molecules are sequestered in planetesimals and in the dusty-icy
matrices that result from collisions of such objects, then dust and CO
production rates should be related.  As many authors have shown, for
debris disks as dusty as those at 49 Cet and HD 21997, collisional
cascade followed by radiative blowout of small dust grains is a far
faster dust loss mechanism than is Poynting-Robertson drag.  

Given the plausibility of a direct or indirect connection between dust and CO production rates (as considered above), 
it is important to understand both rates.  Unfortunately, in two of the best studied debris disks, 
estimates of dust production rates are wildly discrepant -- for Vega (Su et al 2005; M\"uller et al 2010) and for
Fomalhaut (Acke et al 2012; Krivov 2012, personal communication and below). 
We therefore consider collisional cascade rates in some detail.

\vskip 0.2in

\subsection{Dust Loss Rates Via a Collisional Cascade At 49 Cet and HD 21997} 

One may estimate a minimum dust mass required to block a given fraction of the light of a star (e.g., Chen \& Jura 2001); we denote this fraction as 
$\tau$ = L$_{IR}$/L$_{bol}$ (see the third column in Table 5).   For conventional dust grain size 
distributions, such as in a collisional cascade, the dominant particles contributing to $\tau$ have radii somewhat larger than the blowout radius (Chen \& Jura 2001).
The required dust production rate in a steady-state must balance the most rapid relevant loss rate of these dominant particles; in a collisional cascade this loss rate is set
by destructive collisions of small particles that give rise to even smaller particles that can then be radiatively blown out of the debris disk.
$\tau$ is an easily observable quantity and is equivalent to radial optical depth through a spherically symmetric 
dust shell.    

In Appendix B we derive an expression, as a function of $\tau$, for collision times of dust particles with sizes 
slightly above the blowout size.  We consider a cylindrical ring of radius R, width in the radial direction of $\Delta$R, and full vertical height h. 
Then the collision time is

\vskip 0.2in

\begin{equation}
$t$_{coll}$ = P$\Delta$R/24$\tau$R$	
\end{equation}			
			
\vskip 0.2in

\noindent where P is orbital period.  

The dust loss rates at Vega and at Fomalhaut estimated by Su et al (2005) and Acke et al (2012), respectively, are much larger than one calculates 
from  equa.1 or in the collisional cascade model proposed by M\"uller et al (2010).  We consider these discrepancies in detail in Appendix C.   For reasons given there, and because the Su et al and Acke et al papers lack a description of a physical mechanism that can generate small dust particles at the rates 
required in their models, we assume a conventional collisional cascade to produce small particles at a rate given by equa.1.  With this expression and the one that follows, we can calculate the rate of dust grain production/loss for 49 Cet and HD 21997.

As shown in detail by Melis et al (2012), the mass loss rate in a collisional cascade can be written as a function of stellar and dust properties,

\vskip 0.2in

\begin{equation}
$dM/dt = C$\tau$$^2$R$_*$$^{2.5}$T$_*$$^5$M$_*$$^{-0.5}$/T$_{dust}$$
\end{equation}			
			
\vskip 0.2in

\noindent where R$_*$, T$_*$, and M$_*$ are in Solar units and T$_{dust}$ is in K. 
The value of the coefficient C depends on the ratio $\Delta$R/R in equa.1.  For $\Delta$R/R = 0.5, appropriate for 49 Cet and for HD 21997 (Table 5),  C = 1.75 x 10$^{20}$ when 
dM/dt is in g s$^{-1}$.  We take R$_*$, T$_*$, $\tau$, and T$_{dust}$ from Table 2 in Rhee et al (2007), and assume M$_*$ = 2.7 and 2.4 for 49 Cet and HD 21997, respectively.
The derived loss rate, in g s$^{-1}$, for small dust particles is listed in Table 5 for these two stars as well as for Vega and Fomalhaut.  The Table 5 dust loss rates for 49 Cet and HD 21997 are similar to dust production rates derived by Kenyon \& Bromley (2010) in their model for debris disks around 30 Myr old A-type stars (e.g., see their Figs. 13 \& 14).

\vskip 0.2in

\subsection{Ratio of Dust and Gas Production Rates}

In a steady-state model for the CO currently in orbit around HD 21997, Moor et al (2011) deduce that 140 Earth masses of planetesimals in the debris disk would have to have been destroyed in the past 20 Myr.  This contrasts dramatically 
with the M\"uller et al (2010) model for Vega where a collisional cascade operating for the past $\sim$340 Myr 
would have destroyed only a few 
Earth masses of solids.  Moor et al thus favor a situation where the CO is currently experiencing a temporary high production rate.

The mass loss rate in dust at HD 21997 (Table 5) presents another potential problem for any steady state model for 
generating the observed CO.  Moor et al
derive a minimum CO production rate of 10$^{14}$ g s$^{-1}$.  
This is an order of magnitude faster than the dust production rate (Table 5).  Suppose the dust is composed of a 50-50 mixture by mass of water ice versus silicates plus other solids (e.g., Section 3.2 in Acke et al 2012).    Optimistically taking the CO and CO$_2$ mass fractions in a typical comet to each be 20\% that of water ice (Mumma \& Charnley 2011), and assuming that photodissociation of CO$_2$ produces a
CO molecule, one then obtains a ratio, CO/dust, equal to $\sim$0.2 by mass.   Then the CO production rate would seem to be of order a factor of 50 too fast relative to that of the dust, at least in the steady-state. 

These assumed CO and CO$_2$ mass fractions are near the upper end of those measured in solar 
system comets (Mumma \& Charnley 2011).   However, solar system observations sample only the outer layers of comets that are much older and have been heated to much higher temperatures than young comets at 49 Cet and HD 21997.
Thus, it would not be surprising if the latter are, on average, more CO- and CO$_2$-rich than solar system comets. 
The discrepancy between the CO and dust production rates in the steady state could be reduced somewhat should CO and CO$_2$ carry an even larger 
fraction of the mass of a typical youthful comet than we have assumed.

\subsection{Excitation of CO Rotational Levels}

In Section 4.4 we present a colliding comet model to account for the CO observed at 49 Cet and HD 21997.  Should outgassing from comets be responsible for the CO, then the H$_2$/CO ratio in the debris disks is essentially unconstrained.  In particular, there is no reason to expect the large ratios by number, e.g. 10$^4$ or 10$^3$, conventionally assumed in models of
CO excitation (e.g, Hughes et al 2008; Moor et al 2011).  Then excitation of CO rotational levels may well be due to something other than collisions with H$_2$.  One plausible alternative could be collisions of CO with electrons.  Following photodissociation of CO, the C will be photoionized and a source of electrons.  Provided the C+ ions remain in the same volume as the CO molecules for a sufficient length of time, then these electrons can excite the CO rotational ladder.

Dickinson et al (1977) give expressions for calculation of rate coefficients for electron excitation of rotational levels of linear polar molecules in low density clouds.  We assume an electron temperature of 70 K and consider the J = 2-1 transition of CO.  Then the collision time (sec) for excitation of the J = 2 level is 1.4 x  10$^8$/n$_e$, where n$_e$ is the number of electrons per cm$^{-3}$.  For the CO 2-1 transition 1/A $\sim$10$^6$ sec.  So, the J = 2 level can be significantly excited in regions with n$_e$ $\sim$100 cm$^{-3}$.

The electron densities are related to the CO volume densities and the lifetime of C+ and electrons in the volume where the CO resides.  Given the substantial uncertainties in the volumes occupied by the observed CO molecules at 49 Cet and HD 21997, reliable estimation of electron densities will have to await mapping of CO with ALMA.  That said, with CO volumes based on values for $\Delta$R, R and h listed in Table 5, provided that CO and C+ abundances are comparable in regions where CO resides, then electrons would be a viable collision partner for excitation of CO in orbit around 49 Cet and HD 21997.

Independent of whether or not electrons dominate excitation of CO rotational levels, because H$_2$ helps to shield CO against photodissociation, if H$_2$ abundances are much lower than assumed in conventional models, then this could have substantial implications for the CO production rates required to match the observations. 

\subsection{A Comet Model For CO Around 49 Cet and HD 21997} 

Two principal constraints on a model that seeks to explain the CO abundances
at 49 Cet and HD 21997 are (1) the rapid rate of CO production necessitated by
its short ($\sim$ 500 yr) lifetime against photodissociation (e.g., Moor et al
2011), and (2) the relatively rapid rate of CO production compared to dust
production (Section 4.2).

The model outlined below involves collisions among the great numbers of comets
that likely orbit 49 Cet and HD 21997.  Because these two stars rank among the
dustiest A-type stars in the solar vicinity (Section 4 and Table 5) it is not
unreasonable to assume that their debris disks are quite massive relative to
those that orbit most other stars.  In this sense 49 Cet and HD 21997 are very
dusty stars; but in comparison to the amount of CO present, there is relatively
little dust.

Schlichting \& Sari (2011) present a model of the young Sun's Kuiper Belt (KB).
They argue that the total KB mass was dominated during its first 70 Myr by
comet-size planetesimals and that the KB never possessed many more Pluto-size
objects than it does now.  They take the radius R, radial width $\Delta$R, and
total mass of the KB to be 40 AU, 10 AU, and 40 Earth masses, respectively.

We envision the debris regions at 49 Cet and HD 21997 to have R = 100 AU and
($\Delta$R)/R = 0.5 (Table 5).  Neither R nor $\Delta$R are particularly well
constrained by existing data for these stars; hopefully ALMA will substantially
improve the situation in the future.  For the Sun's young Kuiper Belt region and
for 49 Cet and HD 21997 we assume a disk surface density proportional to R$^{-1.5}$
(Lissauer 1987) and a scale height proportional to R.  Then, scaling from the
KB, the total mass of comets that orbit 49 Cet and HD 21997 is equal to
$\sim$400 Earth masses.  This estimate assumes that at a given value of R, say 40 AU
for example, youthful A-type stars will have about the same surface density of comets as did 
the youthful KB.

With CO and CO$_2$ cometary mass fractions as assumed in Section 4.2, 400 Earth masses of
comets corresponds to a CO plus CO$_2$ mass of 80 Earth masses.  With a canonical H$_2$/CO ratio
by number of 10$^4$, a protoplanetary disk of at least 0.2 M$_{\odot}$ is implied.  In a sample of
86 pre-main-sequence stars in the Taurus-Auriga dark clouds, Beckwith et al (1990) deduced the presence
of a few dusty protoplanetary disks with masses as large as a solar mass.  The mass of the typical star in the Beckwith
et al sample is a few times smaller than the masses of 49 Cet and HD 21997, who, as noted above, 
rank among the dustiest main sequence stars known. 

We consider each comet to have a radius of 1 km and a
density of 2 g cm$^{-3}$, such that a typical comet has a mass of 10$^{16}$ g.  
There are thus 2.4 x 10$^{14}$ comets in orbit around 49 Cet and HD 21997.

Sanford \& Alamendola (1990) and Tielens et al. (1991) consider sublimation temperatures for pure CO \& CO$_2$ ices
and also CO \& CO$_2$ ices trapped in an H$_2$O ice matrix; these temperatures are all $\leq$60 K.  For typical 
collision velocities of 100s of m s$^{-1}$ (Appendix B), the bulk of the shattered comets will be heated well above these temperatures.
In addition, "Most kilometer-sized asteroids are likely rubble-piles. Many comets may also be strengthless or nearly strengthless bodies, their fragility demonstrated when
they break up far from perihelion for no obvious reason" (see Movshovitz et al 2012 and references therein).  Given the high CO and CO$_2$ mass fractions assumed in our model, following a cometary collision these molecules should have little trouble 
escaping as vapors.  At the same time, less volatile materials can remain as solids (e.g., Tielens et al 1994; Czechowski \& Mann 2007).

We estimate the characteristic time for comet collisions using equa. B6.  The
number of comets per unit volume (N) is equal to 2.4 x 10$^{14}$ divided by the volume of
the debris ring (2$\pi$R$\Delta$Rh).  This yields a collision time of

\begin{equation}
$t$_{coll}$ = 2R$\Delta$RP/$\pi$(10$^{25}$)$
\end{equation}

\noindent where R and $\Delta$R are in cm.  With a period of 650 yr at 100 AU around these A-type stars, t$_{coll}$ = 5 x 
10$^7$ yr. With 2.4 x 10$^{14}$ comets, there are 5 x 10$^6$ collisions/yr, or one every 6 seconds!

As in Section 4.2, we assume that 50\% of the mass of a typical comet is water
ice, that CO and CO$_2$ each comprise 20\% as much mass as water ice, and that
photodissociation of CO$_2$ yields a CO molecule.  When two comets collide, we
assume that 50\% of the total mass becomes debris consisting of CO, CO$_2$, and solid
material composed primarily of equal quantities by mass of water ice and silicates.  With these
assumptions, CO is produced at a rate $\sim$10$^{22}$g yr$^{-1}$.  To reproduce 
the observed CO mass of $\sim$3.5 x 10$^{-4}$ Earth masses at HD 21997, Moor et 
al (2011) require a CO production rate of 4 x 10$^{21}$g yr$^{-1}$.  

Hughes et al (2008) do not give a CO production rate appropriate to their model
for 49 Cet, but the CO mass at 49 Cet is about 7 times that at HD 21997 or about the mass
of Pluto (A.M. Hughes, private comm. 2012).  While this is a lot of CO, it is
only a tiny fraction of the total mass of comets in the debris disk.  

As noted in Section 4.2, the CO production rate by mass at HD 21997 is a
factor of 10 faster than the dust production rate given in Table 5 via a
collisional cascade.  If we assume that the CO lifetime against
photodissociation at 49 Cet is the same as that at HD 21997 (500 years), then
for 49 Cet the ratio of CO to dust production rate is $\sim$20.

In our model the fraction of mass released via comet collisions in the form
of CO and CO$_2$ is only $\sim$20\% of the mass released in solid material that
will eventually, via a collisional cascade, become the 10 $\mu$m size dust
grains considered in Appendix C.
The key to why the CO-to-dust production rates are skewed so heavily in favor 
of CO is probably related to the much shorter time it takes to release large 
amounts of CO following a disruptive impact compared to the time to whittle 
a large chunk of solid debris down to 10 $\mu$m-size particles.   With equa.1, the 
collision time at HD 21997 for dust particles with sizes a bit above the 
blowout radius is 2.6 x 10$^4$ yr.  For typical solid chunks coming off a comet 
collision, the ensuing collision times could easily be millions of years.  

In the model of Schlichting \& Sari (2011), dynamical stirring of the young KB
increases as a function of time -- as large bodies grow in the debris belt, they
stir up the smaller objects (the comet-size objects we are considering here).  A similar picture of the 
early evolution of debris disks around youthful A-type stars was developed earlier by Kenyon
\& Bromley (2008, 2010).
Thus, 49 Cet and HD 21997 may be in a dynamically active phase where comet
collisions are frequent, but where the breakdown of large bodies of solid
debris into small dust particles has not yet had time to operate to conclusion.  
The spirit of such a model is not one that relies on an unusual,
catastrophic, transient event.  Rather the two stars are passing through a
normal era in the evolution of comet-dominated debris belts.  Given the 
discussions in Schlichting \& Sari (2011) and Kenyon \& Bromley (2010), it would not be unreasonable for that 
era to occur at an age of $\sim$30 Myr (see, e.g.,, the 1 km panel in Figure 23 in Kenyon \& Bromley 2010).

\section{Conclusions}

We present a massive (400 Earth mass) comet-cloud model to explain the large quantities of carbon monoxide gas seen at
the 30-40 Myr old, A-type, stars HD 21997 and 49 Ceti.  Because CO is rapidly photodissociated in the stellar 
and interstellar radiation fields, it must be produced rapidly.  We calculate that this
production rate is an order of magnitude faster than the rate of production of dust particles
via a model of collisional cascade in the steady state.  The implications of this ratio
are: (1) young, pristine comets are likely to be richer in CO and perhaps CO$_2$ than typically
observed solar system comets, and (2) the destruction of the youthful comets around HD 21997 and
49 Cet are not likely to have been in a steady state situation over periods of time exceeding millions of years.
If dynamical activity in debris disks is ramping up at the age of these stars, for example through the build-up of
large bodies as in the model of the young Kuiper Belt proposed by Schlichting \& Sari (2011) or the model of debris disks
around A-type stars by Kenyon \& Bromley (2010), then the production
of small dust particles may lag well behind the rate of CO outgassing.

The primary driving force for CO outgassing is likely to be collisions among the 100 trillion comets that orbit 49 Cet and HD 21997.
Such a model for these systems is consistent with origin in a protoplanetary
disk of initial mass of order a few tenths of a solar mass and suggests that comets form very early in the 
history of such disks before there is sufficient time to convert much CO to CH$_4$.  If comet collisions provide the observed CO,
then the H$_2$/CO ratio is unconstrained and may not be large.  If so, then collisional excitation of the CO rotational levels may be
dominated by something other than H$_2$, perhaps electrons.

Acke et al (2012) have proposed a rapid comet-destruction model for the dusty debris disk that orbits Fomalhaut
that, at first glance, might seem similar to our model for HD 21997 and 49 Ceti.   But, in fact, the two models are 
very different.  As we note in Appendix C, the Acke et al dust loss rate is 20 times faster than the rate
we calculate via a conventional collisional cascade.  They do not specify a physical mechanism that can convert
comets so rapidly into dust.  Their rapid dust loss rate at Fomalhaut is compelled by their quoted very short lifetime (1700 yr) for dust
grains near the blowout size.  But a detailed model of the similar debris disk at Vega by M\"uller et al (2010 and as outlined 
in our Appendix C) casts doubt on the validity of such a short lifetime.  Additional observations and modeling of 
the Fomalhaut debris disk are warranted.   In particular, a deep search for CO at Fomalhaut would be worthwhile; because it is 
the presence of so much CO that compels the rapid comet destruction rate in our model, while combination of observations of CO and of dust
serve to constrain the model.
 
 At the risk of being obvious, dusty young stars will be excellent targets for ALMA mapping of 
 dust and CO gas, should the latter be detectable.  It should be possible to map the shape 
 and mass of youthful Kuiper Belt analogs and to clarify the composition of young comets.
 
\vskip 0.2in

We thank Meredith Hughes, Alexander Krivov, and Hilke Schlichting for taking time to 
clarify some aspects of their papers, David Jewitt and 
Michael Jura for helpful comments, and the referee for useful suggestions.
This research was funded in part by NASA grants to UCLA and the University of Georgia.

\appendix

\section{IRAM 30 m CO Observations}

Table 4 lists a set of upper limits on peak intensities ($3\sigma_T$) and integrated intensities in the $^{12}$CO(2--1) line ($5\sigma_I$) as measured with the IRAM 30 m telescope in 2009 August for a sample of dusty stars within $\sim$100 pc of Earth that includes HD 21997. These upper limits supercede those reported in columns 8--11 in Table 1 of Kastner et al.\ (2010). The upper limits on integrated line intensity are computed as in \S 2.1 of Kastner et al. (2010) for spectrometer channel widths of $\delta v = 0.8$ km s$^{-1}$; but now for an assumed linewidth of $\Delta v = 5$ km s$^{-1}$, whereas they assumed a linewidth of $\Delta v = 3$ km s$^{-1}$. The equivalent line intensities in Jy km s$^{-1}$ are then obtained assuming a conversion factor of 7.85 Jy K$^{-1}$ for the 30 m telescope at 230 GHz.

The listed values of $3\sigma_T$ were obtained from the rms of the residuals to a fit to the baseline of each  spectrum over the range $-5$ km s$^{-1}$ to $+5$ km s$^{-1}$; these values differ only slightly from those reported in Kastner et al. (2010), which were based on fits over a somewhat longer baseline.  The conversion factor employed here differs from that adopted by Kastner et al. (2010); this factor was applied erroneously by those authors.  Note that the $^{12}$CO(2--1) upper limit for HD 21997 in Table 4, $5\sigma_I = 3.1$  Jy km s$^{-1}$, is consistent with the integrated line intensity of $2.29\pm0.47$ Jy km s$^{-1}$ reported in Moor et al.\ (2011).  Likewise the upper limit on CO mass for HD 21997 in Table 4 is consistent with their measured CO mass.

\section{Collisional Cascade Rates}

The easily observable quantity is radial optical depth through a spherically symmetric 
dust shell, $\tau$ = L$_{IR}$/L$_{bol}$, but collision rates of disk particles are more directly 
related to what has been called "face-on fractional surface density $\sigma$(r)" (Backman \& Paresce 1993) or "full 
vertical optical thickness" (= $\tau$$_\perp$, Artymowicz \& Clampin 1997) of disk-like structures.  The expressions given by these sets of authors are similar and we begin with equation 8 from Artymowicz \& Clampin (1997):

\vskip 0.2in

\begin{equation}
$t$_{coll}$ = P/12($\tau$$_\perp$)$
\end{equation}

\vskip 0.2in

\noindent where P is orbital period.
We consider a cylindrical ring of radius R, width in the radial direction of $\Delta$R, and full vertical height h.  Then the ring (disk) optical depth in the radial direction, $\tau$$_r$,  is

\vskip 0.2in

\begin{equation}
$$\tau$$_r$ = 2R$\tau$/h$
\end{equation}		

\vskip 0.2in

\noindent  Relating $\tau$$_r$ and $\tau$$_\perp$ gives

\vskip 0.2in

\begin{equation}
$$\tau$$_\perp$ = $\tau$$_r$ (h/$\Delta$R) = 2R$\tau$/$\Delta$R$
\end{equation}			
			
\vskip 0.2in

\noindent or with equa. B1,

\vskip 0.2in

\begin{equation}
$t$_{coll}$ = P$\Delta$R/24$\tau$R$	
\end{equation}			
			
\vskip 0.2in

\noindent When P is expressed in terms of orbital radius to the 1.5 power and stellar mass to the -0.5 power (as per Kepler and Newton), then this 
expression is in good agreement with equa. (25) of Wyatt et al (2007).  Finally, with $\Delta$R/R = 0.3 from Table 5, we 
obtain a typical dust particle collision time

\vskip 0.2in

\begin{equation}			
$t$_{coll}$ = P/80$\tau$$
\end{equation}

\vskip 0.2in

\noindent with a variation of perhaps a factor of $\pm$2 due to differing dust ring, in-plane, extents.

One can also employ kinetic theory to independently derive an expression that relates t$_{coll}$ to $\tau$, but without
use of equa. B1.  Collision times are equal to the mean free path (mfp) between dust particle 
collisions divided by the relative particle velocities.   The relative velocity for typical particles in the ring is the orbital velocity 
times h/2R.  So the collision time is

\vskip 0.2in

\begin{equation}
$t$_{coll}$ = (mfp)P/$\pi$h$			
\end{equation}			
			
\vskip 0.2in

\noindent The mfp = ($\sigma$N)$^{-1}$ = (4$\pi$a$^2$N)$^{-1}$, where "a" is dust grain radius and N is the number of grains per unit volume.  Then,

\vskip 0.2in

\begin{equation}
$$\tau$ = N$\pi$a$^2$ (2$\pi$R$\Delta$Rh)/4$\pi$R$^2$$
\end{equation}
			
\vskip 0.2in

\noindent and

\vskip 0.2in

\begin{equation}
$mfp = $\Delta$Rh/8$\tau$R$
\end{equation}			
			
\vskip 0.2in

\noindent thus, 

\vskip 0.2in

\begin{equation}
$t$_{coll}$ = P$\Delta$R/8$\pi$$\tau$R$
\end{equation}

\vskip 0.2in

\noindent in agreement with equa. B4.

Artymowicz \& Clampin (1997) use kinetic theory to derive a collisional rate a few times faster than that 
given in equations B4 and B9, but then choose to adopt the rate given in equa. B4; we do the same.

\vskip 0.2in

\section{Dust Loss Rates At Vega and Fomalhaut}

Based on Spitzer Space Telescope observations of Vega, Su et al (2005) calculated a large
dust production rate, $\sim$6 x 10$^{14}$ g s$^{-1}$, and 
concluded that this rate could not have been sustained over Vega's 350 Myr lifetime.   M\"uller et al (2010) 
reinterpreted the Spitzer data and derived a dust production/loss rate via collisional cascade 
two orders of magnitude smaller than that of Su et al.  In the M\"uller et al model the mass in dust grains with radii 
$\sim$10 $\mu$m -- that is, slightly above the blowout size -- is $\sim$7 x 10$^{-4}$ Earth masses.  One can calculate a
similar dust mass from an estimate of the dust surface area  
required to block 2 x 10$^{-5}$ of the light from Vega (e.g., equa. 4 in Chen \& Jura 2001).  In their elaborate Vega model, M\"uller et al (2010) and A. Krivov (private comm. 2012) derive a collisional lifetime for these small particles of $\sim$3 x 10$^5$ yr.
Together, equa. B4 and Table 5 yield a collision time of 5 x 10$^5$ yr, and thus a dust mass loss rate via a collisional cascade of $\sim$3 x 10$^{11}$ g s$^{-1}$.  We note in passing, there is a typo in the value for $\tau$$_\perp$ = 8.3 x 10$^{-4}$
given in Section 6.8 of M\"uller et al (2010).  In fact, $\tau$$_\perp$ in their model is 8.3 x 10$^{-5}$ (A. Krivov, private comm. 2012), 
in good agreement with the value of 10$^{-4}$ we derive from equa. B3 and Table 5.

Based on Herschel Space Observatory images, Acke et al (2012), derived mass loss rates in 
dust grains for Fomalhaut. Given that their methods are similar to those used by Su et al (2005) to analyze 
Vega, it is surprising that Acke et al cite neither Su et al, nor M\"uller et al (2010).   Like Su et al, Acke et al derive small grain lifetimes of
$\sim$1000 yr, that is, much shorter than can be achieved via grain production with a conventional collisional cascade.  
Thus, if the Acke et al mass loss 
rate of $\sim$7 x 10$^{13}$ g s$^{-1}$ were appropriate 
for Fomalhaut, then dust grain production would likely be due to something other than a collisional cascade (see paragraph that follows);
Acke et al (2012) suggest a quasi-steady state model of colliding comets over a 200 Myr lifetime of Fomalhaut.  Should Fomalhaut's age instead be
$\sim$440 Myr, as seems likely (Mamajek 2012), then the Acke et al picture would become even more extreme.

Blockage of  8 x 10$^{-5}$ of the light from Fomalhaut with 10 $\mu$m radius grains of density 2 g cm$^{-3}$, requires a 
mass of $\sim$10$^{25}$ g.  This may be compared with the mass of 3 x 10$^{24}$ g of comparable size grains with, 
apparently, somewhat lower density in the model of Acke et al (2012).  With equa. B4, we find a characteristic  
dust production timescale of 8 x 10$^4$ yr and a mass loss rate of 4 x 10$^{12}$ g s$^{-1}$ via collisional cascade.  
This mass loss rate is $\sim$20 times slower 
than the Acke et al mass loss rate.  Since the mass and luminosity of Fomalhaut are similar to that of $\zeta$ Lep (Rhee et al 2007;
Chen \& Jura 2001), compact grains of radii only a few microns would be stable against blowout, thus potentially increasing 
the discrepancy between the quoted Acke et al mass loss rate and that estimated from a collisional cascade.   It may be possible
to match the spectral energy distribution and images of Fomalhaut with inclusion of such small, bound, grains (A. Krivov, private comm. 2012).
Should a model similar to that of M\"uller et al (2010) apply to Fomalhaut as well as to Vega,
then a normal collisional cascade might account for the dust distribution at Fomalhaut and invocation of an extreme
colliding-comet model might be unnecessary.  In particular, use of a single power-law size distribution for both blow-out and bound grains 
is unphysical (A. Krivov, private comm. 2012; M\"uller et al 2010) and additional modeling of the data is warranted.

As noted in the main text, it is one thing to have comets collide, but converting the resulting debris into small dust particles is something else again.  
The Su et al and Acke et al papers lack a description of a physical mechanism that can generate small dust particles at the rates 
required in their models.  Therefore, in agreement with M\"uller et al (2010), we regard a conventional collisional cascade as the most plausible dust production mechanism. 

\noappendix

\section*{References}
\begin{harvard}

\item[Acke, B., Min, M., Dominik, C. et al 2012, arXiv:1204.5037]
\item[Artymowicz, P. \& Clampin, M. 1997, ApJ 490, 863]
\item[Backman, D. \& Paresce, F. 1993, in Protostars and Planets III, ed. V. Mannings et al. (Tucson: Univ.]
\indent Arizona Press), 1253
\item[Beckwith, S., Sargent, A., Chini, R. \& Guesten, R. 1990, AJ 99, 924]
\item[Boley, A., Payne, M., Corder, S., et al. 2012, ApJL 750, L21]
\item[Chen, C. \& Jura, M. 2001, ApJ 560, L171]
\item[Czechowski, A. \& Mann, I. 2007, ApJ 660, 1541]
\item[Dent, W. R.F., Greaves, J. S. \& Coulson, I. M. 2005, MNRAS 359, 663]
\item[Dickinson, A., Phillips, T., Goldsmith, P., Percival, I. \& Richards, D. 1977, A\&A 54, 645]
\item[Fekel, F. C., Tomkin, J. \& Williamson, M. H. 2009 AJ 137, 3900]
\item[Gontcharov, G. 2006, Astron. Lett. 32, 759] 
\item[Hughes, A., Wilner, D., Kamp, I. \& Hogerheude, M. 2008, ApJ 681, 626]
\item[Jura, M., Malkan, M., White, R. et al 1998, ApJ 505, 897]
\item[Kastner, J. H, Hily-Blant, P., Sacco, G. G., Forveille, T. \& Zuckerman, B. 2010, ApJ 723, L248]
\item[Kenyon, S. \& Bromley, B. 2008, ApJS 179, 451]
\item[Kenyon, S. \& Bromley, B. 2010, ApJS 188, 242]
\item[Kharchenko, N.  \& Roeser, S. 2009, VizieR On-line Data Catalog]
\item[Lissauer, J. 1987, Icarus 69, 249]
\item[Mamajek, E. 2012, ApJ in press (arXiv:1206.6353)]
\item[Melis, C., Zuckerman, B., Rhee, J. et al 2012, ApJ submitted]
\item[M\"uller, S., L\"ohne, T. \& Krivov, A. 2010, ApJ 708, 1728]
\item[Moor, A., Abraham, P., Derekas, A., Kiss, Cs., Kiss, L., Apai, D., Grady, C. \& Henning, Th.]
\indent 2006, ApJ 644, 525
\item[Moor, A., Abraham, P., Juhasz, A. et al 2011, ApJ 740, L7]
\item[Movshovitz, N., Asphaug, E. \& Korycansky, D. 2012, arXiv:1207.3386] 
\item[Mumma, M. \& Charnley, S. 2011, ARAA 49, 417]
\item[Ortega, V., de la Reza, R., Jilinski, E. \& Bazzenella, B. 2002, ApJ 575, L75] 
\item[Rhee, J., Song, I., Zuckerman, B. \& McElwain, M. 2007, ApJ 660, 1556]
\item[Roberge, A. \& Weinberger, A. 2008, ApJ 676, 509] 
\item[Rodriguez, D. R., Kastner, J. H., Wilner, D. \& Qi, C. 2010, ApJ 720, 1684]
\item[Sanford, S. \& Alamendola, L. 1990, Icarus 87, 188]
\item[Schlichting, H.  \& Sari, R. 2011, ApJ 728, 68]
\item[Su, K., Rieke, G., Misselt, K. et al. 2005, ApJ 628, 487]
\item[Tielens, A., McKee, C., Seab, C. \& Hollenbach, D. 1994, ApJ 431, 321]
\item[Tielens, A., Tokunaga, A., Geballe, T. \& Bass, F. 1991, ApJ 381, 191]
\item[Torres, C., Quast, G., Melo, C. \& Sterzik, M. 2008, "Handbook of Star Forming Regions,] 
\indent Volume II: The Southern Sky" ASP Monograph Publ., Vol. 5., ed. Bo Reipurth, p.757
\item[van Leeuwen, F. 2007, A \& A 474, 653]
\item[Vigan, A., Patience, J., Marois, C. et al. 2012, ApJ in press (arXiv:1206.4048)]
\item[Wahhaj, Z., Koerner, D. W. \& Sargent, A. I. 2007, ApJ 661, 368]
\item[Wyatt, M., Smith, R., Greaves, J. et al 2007, ApJ 658, 569]
\item[Zuckerman, B. 2001, ARAA 39, 549]
\item[Zuckerman, B., Forveille, T. \& Kastner, J. H. 1995, Nature 373, 494]
\item[Zuckerman, B., Melis, C., Rhee, J., Schneider, A \& Song, I. 2012, ApJ 752, 58]
\item[Zuckerman, B., Rhee, J. H., Song, I. \& Bessell, M. S. 2011, ApJ 732, 61]
\item[Zuckerman, B. \& Song, I. 2004, ARAA 42, 685]
\item[Zuckerman, B., Song, I., Bessell, M. \& Webb, R. 2001, ApJ 562, L87]
\item[Zuckerman, B. \& Webb, R. 2000, ApJ 535, 959]
 
\end{harvard}

\clearpage

\begin{landscape}
\begin{table}
\caption{Proposed Argus Association Members}
\begin{tabular}{@{}lcccccccccc}
\br
HIP& HD& R.A.& Decl.& Spec.& V& B-V& Dist.& Rad. Vel.& (U,V,W)& UVW error\\
& & (h/m)& (deg)& Type& (mag)& (mag)&  (pc)& (km s$^{-1}$)& (km s$^{-1}$)& (km s$^{-1}$)\\
\br
7345& 9672& 01 34& -15& A1& 5.6& 0.07& 59&  10.3$\pm$0.7& -22.9,-16.4,-5.4& 0.4,0.3,0.7\\
23585& 32296& 05 04& +45& A2& 6.5& 0.22& 61& 19.4$\pm$1.0& -24.4,-11.1,-5.4& 1.0,0.7,0.3\\
51194& 90874& 10 27& -65& A2& 6.0& 0.09& 68& 7.1$\pm$1.0& -22.8,-14.3,-8.8& 0.5,0.6,0.2\\
57013& 101615& 11 41& -43& A0& 5.5& 0.04& 65& 8$\pm$2.5& -20.1,-17.3,-2.9& 0.9,2.2,0.8\\
76736& 138965& 15 40& -70& A1& 6.4& 0.08& 78& -2$\pm$4.3& -19.4,-15.5,-6.3& 3.1,2.9,0.9\\
89925& 168913& 18 20& +29& A5& 5.6& 0.23& 56& -21.9$\pm$2.9& -23.4,-11.6,-2.4& 1.5,2.3,0,9\\
98103& 188728& 19 56& +11& A1& 5.3& 0.01& 67& -28.0$\pm$4.2& -23.2,-18.5,-3.5& 2.6,3.2,0.7\\
 \br
\end{tabular}
\end{table}
\noindent Notes $-$ Input data (R.A., Decl., distance, proper motion) for the UVW calculations in Tables 1 and 2 are from the Hipparcos catalog (van Leeuwen 2007).  Radial velocities and their errors are from Gontcharov (2006), except for HIP 57013 where the velocity and its error are from VizieR.  UVW are defined with respect to the Sun, with U positive toward the Galactic Center, V positive in the direction of Galactic rotation, and W positive toward the North Galactic pole. Additional information regarding the listed stars can be found in Sections 2 and 3. HIP 7345 is 49 Cet.  Torres et al (2008) give a mean UVW for Argus stars of -22.0$\pm$0.3, -14.4$\pm$1.3, -5.0$\pm$1.3 km s$^{-1}$.  Among the listed stars, HIP 98103 appears the most likely to be a non-member (see Table 3).  
\end{landscape}

\clearpage

\begin{landscape}
\begin{table}
\caption{Proposed Tucana/Horologium \& Columba Association Members}
\begin{tabular}{@{}lcccccccccc}
\br
HIP& HD& R.A.& Decl.& Spec.& V& B-V& Dist.& Rad. Vel.& (U,V,W)& UVW error\\
& & (h/m)& (deg)& Type& (mag)& (mag)&  (pc)& (km s$^{-1}$)& (km s$^{-1}$)& (km s$^{-1}$)\\
\br
25280& 35505& 05 24& -16& A0& 5.6& 0.0& 68& 21.0$\pm$1.2 & -11.8,-21.2,-1.7& 0.7,0.6,0.5\\
26395& 37306& 05 37& -11& A2& 6.1& 0.05& 63& 23.0$\pm$1.0 & -13.1,-20.2,-5.8& 0.5,0.4,0.3\\
41081& 71043& 08 22& -52& A0& 5.9& 0.02& 70& 22.5$\pm$2.0 & -10.3,-21.6,-5.9& 0.2,1.1,0.2\\
53771& 95429& 11 00& -51& A3& 6.2& 0.19& 61& 13.9$\pm$1.0 & -12.6,-19.0,-5.3& 0.5,1.0,0.2\\
59394& 105850& 12 11& -23& A1& 5.5& 0.06& 59& 11.0$\pm$4.2 & -10.5,-19.6,-0.3& 1.3,3.1,2.6\\
72408& 130556& 14 48& +21& F0& 7.9& 0.41& 75& -9.5$\pm$1.5 & -11.3,-20.6,-0.5& 0.7,1.0,1.4\\
104365& 201184& 21 08& -21& A0& 5.3& 0.00& 55& -12.0$\pm$3.2& -7.7,-18.9,0.0& 2.2,1.2,2.1\\
\br
\end{tabular}
\end{table}
\noindent Notes $-$  Characteristic mean UVW for Tuc/Hor stars are given in Zuckerman \& Song (2004) as -11, -21, 0 km s$^{-1}$, and in Torres et al. (2008) as -9.9$\pm$1.5, -20.9$\pm$0.8, -1.4$\pm$0.9 km s$^{-1}$.  Torres et al give a mean UVW for Columba stars of -13.2$\pm$1.3, -21.8$\pm$0.8, -5.9$\pm$1.2 km s$^{-1}$.   There is a galaxy $\sim$50 arcsec from HIP 72408 (mostly toward the south) that provides the strong 60 \& 100 $\mu$m IRAS excess emission noted in VizieR.
HIP 72408 is apparently a weak X-ray source in the ROSAT All-Sky Survey Catalog.   HIP 30675 (= HR 2328) may be a Columba star at a distance from Earth of $\sim$125 pc (see Section 3).

\end{landscape}

\clearpage

\begin{table}
\caption{Key for Figure 1}
\begin{tabular}{@{}lcccccc}
\br
HIP& V& B-V& M$_V$ & Moving& Age& Reference\\
& (mag)& (mag)& (mag)& Group& (Myr)&  for M.G.\\
\mr
A-49 Cet& 5.62& 0.07& 1.75& Argus& 40&  1\\
B-2578& 5.07& 0.04& 1.78& Tuc/Hor& 30&  2\\
C-12413& 4.74& 0.06& 1.98& Tuc/Hor& 30&  3\\
D-15353& 6.03& 0.13& 2.33& AB Dor& 70&  3\\
E-16449& 6.38& 0.12& 2.10& Columba& 30& 4 \\
F-23179& 4.93& 0.04& 1.34& Columba& 30*&  3\\
G-25280& 5.65& 0.00& 1.48& Tuc/Hor& 30&  1\\
H-26309& 6.26& 0.15& 2.65& Columba& 30& 3\\
I-26395& 6.10& 0.05& 2.11& Columba& 30& 1\\
J-26966& 5.73& -0.01& 1.35& Columba& 30& 4\\
K-32104& 5.20& 0.06& 2.00& Columba& 30& 3\\
L-41081& 5.89& 0.02& 1.66& Columba& 30& 1\\
M-50191& 3.85& 0.05& 1.39& Argus& 40*& 3\\
N-51194& 6.00& 0.09& 1.84& Argus& 40& 1\\
O-57013& 5.54& 0.04& 1.46& Argus& 40&  1\\
P-57632& 2.14& 0.09& 1.93& Argus& 40& 3\\
 Q-59394& 5.45& 0.06& 1.60& Tuc/Hor& 30&  1\\
 R-76736& 6.42& 0.08& 1.94& Argus& 40&  1\\
 S-95261& 5.03& 0.02& 1.61& $\beta$ Pic& 12&  5\\
 T-98103& 5.28& 0.01& 1.13& Argus& 40*&  1\\
 U-98495& 3.97& -0.03& 1.43& Argus& 40&  3\\
 V-99770& 4.93& 0.15& 1.78& Argus& 40*&  3\\
 W-104365& 5.30& 0.00& 1.60& Tuc/Hor& 30&  1\\
 X-115738& 4.95& 0.04& 1.59& AB Dor& 70&  3\\
 Y-117452& 4.59& 0.00& 1.47& AB Dor& 70&  3\\
 Z-118121& 5.00& 0.06& 1.62& Tuc/Hor& 30&  2\\
\br
\end{tabular}
%\end{table}

\noindent Notes $-$ With the exception of 49 Cet, the letters preceding the HIP designation are those that are plotted on Figure 1.  HD 21997 is star E. Additional information regarding some listed stars can be found in Section 3.   An asterisk in the right hand column indicates stars that appear high (bright) on the figure and thus might be too old to be members of the designated young moving group.  Some problems with assigning A-type stars to moving groups are considered in Sections 2 and 3.  The right hand column
gives the earliest reference where one can find a given star placed into a given moving group:  (1) this paper, (2) Zuckerman \& Webb 2000, (3) Zuckerman et al (2011), (4) Moor et al (2006), (5) Zuckerman et al (2001).
\end{table}

\clearpage

\begin{table}
\caption{IRAM 30 m CO Observations of Nearby, Dusty Stars}
%\label{tbl:nondetections}
%\begin{center}
\begin{tabular}{lcccc}
  \br
  name & 3$\sigma$$_T$ & 3$\sigma$$_I$ & 5$\sigma$$_I$  & M$_{CO}$ \\
  &  (mK) & (mK km s$^{-1}$) & (Jy km s$^{-1}$) & (10$^{-4}$ M$_E$) \\
  \mr
     HD 15745     &   72    &   145    &   1.9& $<$5.0 \\
     HD 21997      & 119   &    240   &    3.1& $<$11.0 \\
     HD 30447     &  139   &    280   &    3.7& $<$14.5 \\
     HD 32297     &   89    &   178   &   2.3& $<$18.5 \\
     HD 38206    &   146   &    295   &    3.9& $<$11.8 \\
     HD 85672     &   64    &   128   &   1.7& $<$9.2 \\
    HD 107146    &    55    &   110    &    1.4& $<$0.8 \\
    HD 131835    &   193   &    388    &     5.1& $<$40.6 \\
    HD 191089     &  124   &    249   &    3.3& $<$5.9 \\
    HD 221853     &   54    &   110    &  1.4& $<$5.0 \\
\br
\end{tabular}
%\end{center}

\noindent Notes $-$ This is a revision of columns 8-11 in Table 1 in Kastner et al (2010).  The listed main beam brightness temperatures (column 2) and the upper limits on integrated line intensity are all only upper limits; 3 $\sigma$ for the second and third columns and 5 $\sigma$ for columns 4 and 5.  See Appendix A for details.  As noted there, the linewidth assumed for calculation of upper limits to integrated line intensity in Table 4 is 
5 km s$^{-1}$, whereas 3 km s$^{-1}$ was assumed in Table 1 in Kastner et al (2010).  M$_E$ is the mass of Earth.
%\footnotesize
%\end{sidewaystable}
\end{table}

\clearpage

\begin{table}
\caption{Disk Parameters}
\begin{tabular}{@{}lcccccccc}
\br
Star &  Age &$\tau$ & M$_{dust}$ & M$_{CO}$ & $\Delta$R/R & h/R & dM/dt & Reference for\\
&  (Myr) & (10$^{-4})$ & (M$_E$) & (10$^{-4}$M$_E$)  & & & (10$^{12}$ g s$^{-1}$) & $\Delta$R \& h\\
\mr
Vega & 350 &  0.2 & 0.008  &  & 0.4 & 0.2 & 0.3  & Muller et al 2010\\
Fomalhaut & 440 & 0.8 & 0.024 &  & 0.13 &  & 4 & Acke et al 2012;\\
& & & & & & & & Boley et al 2012\\
49 Cet & 40 &  8 & 0.31 & 25 & $\sim$0.5 & 0.04 & 45 & Hughes et al 2008\\
HD 21997& 30 & 5 & 0.22 & 3.5 & $\sim$0.5  & 0.2 & 13 & Moor et al 2011\\
\br
\end{tabular}

\noindent Notes $-$ Values of $\tau$ (= L$_{IR}$/L$_{bol}$) are from Rhee et al (2007).  R is disk radius, $\Delta$R is disk width in the radial direction, and h is full vertical height of the disk at radius R.   Artymowicz \& Clampin (1997) give h/R $\sim$0.2 for the disk that orbits $\beta$ Pictoris.  The dust masses, M$_{dust}$, obtained directly from submillimeter measurements (Rhee et al 2007), are in units of Earth masses.  The CO masses are in units of 10$^{-4}$ Earth masses.  dM/dt is the production/loss rate of dust grains with radii slightly larger than the blow-out radius in a quasi-steady state, collisional cascade, model  (calculated in Appendix C for Vega and Fomalhaut and in Section 4.1 for 49 Cet and HD 21997).  The listed age for Fomalhaut is from a recent paper by Mamajek (2012). 
\end{table}

\clearpage

\begin{figure}
\includegraphics[width=140mm]{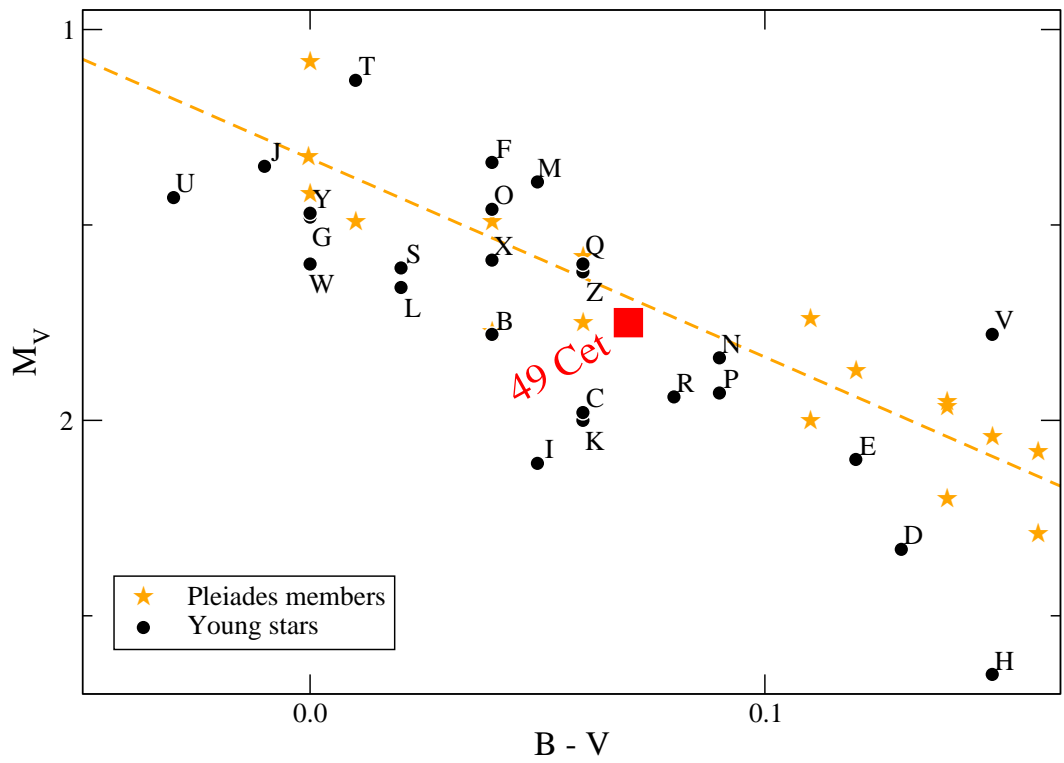}
\caption{\label{figure1} Color-magnitude diagram of stars listed in Table 3.  Gold stars are members of the Pleiades cluster that have been dereddened and cleaned of 
binary stars (from Vigan et al 2012).  The dashed line is an empirical linear fit to Pleiades stars with B-V between 0 and 0.3 magnitudes (from Vigan et al 2012).}
\end{figure}

\clearpage

\begin{figure}
{\centering \begin{tabular}{cc}
\includegraphics[scale=0.4]{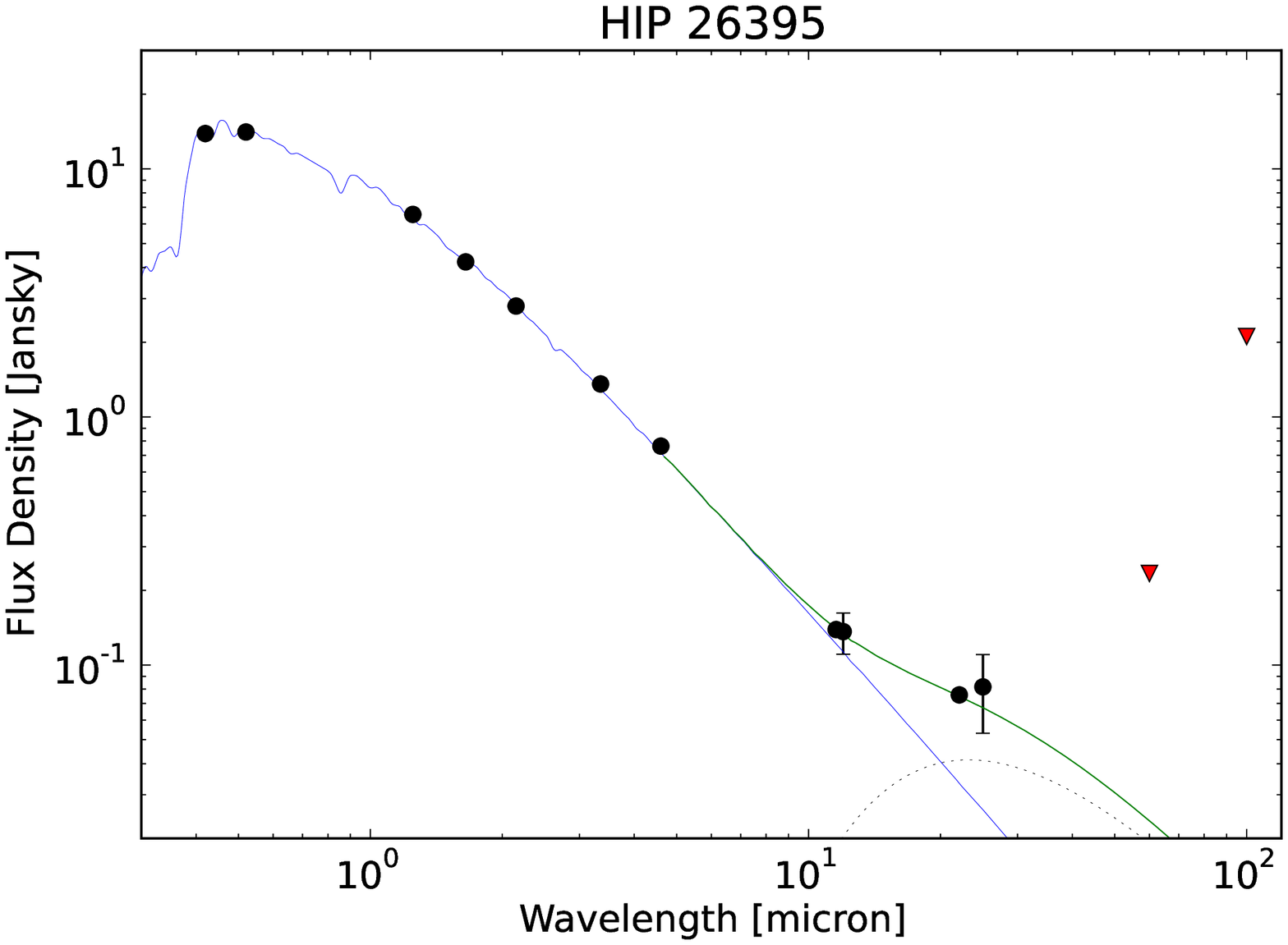} &
\includegraphics[scale=0.4]{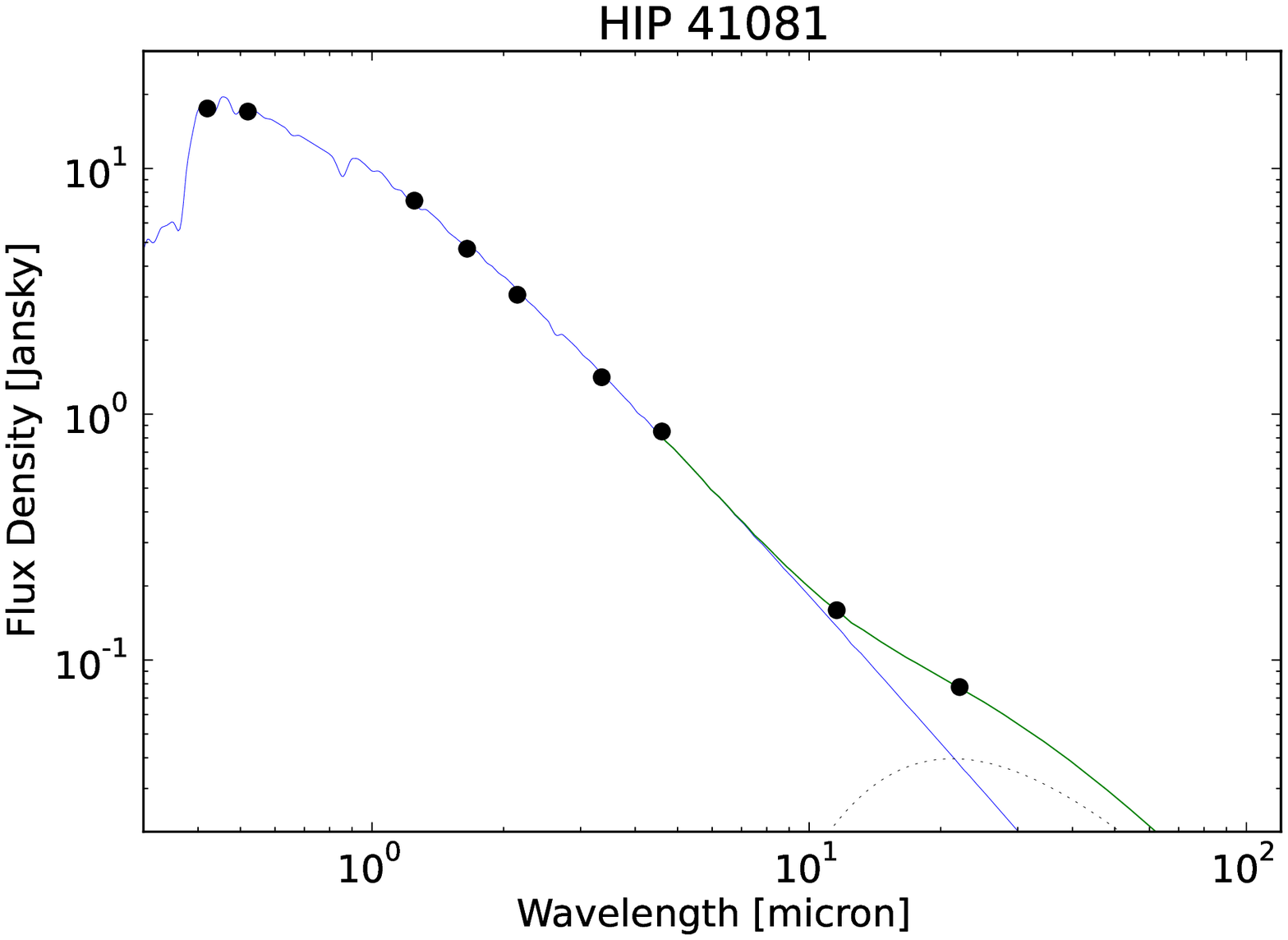} \\
\includegraphics[scale=0.4]{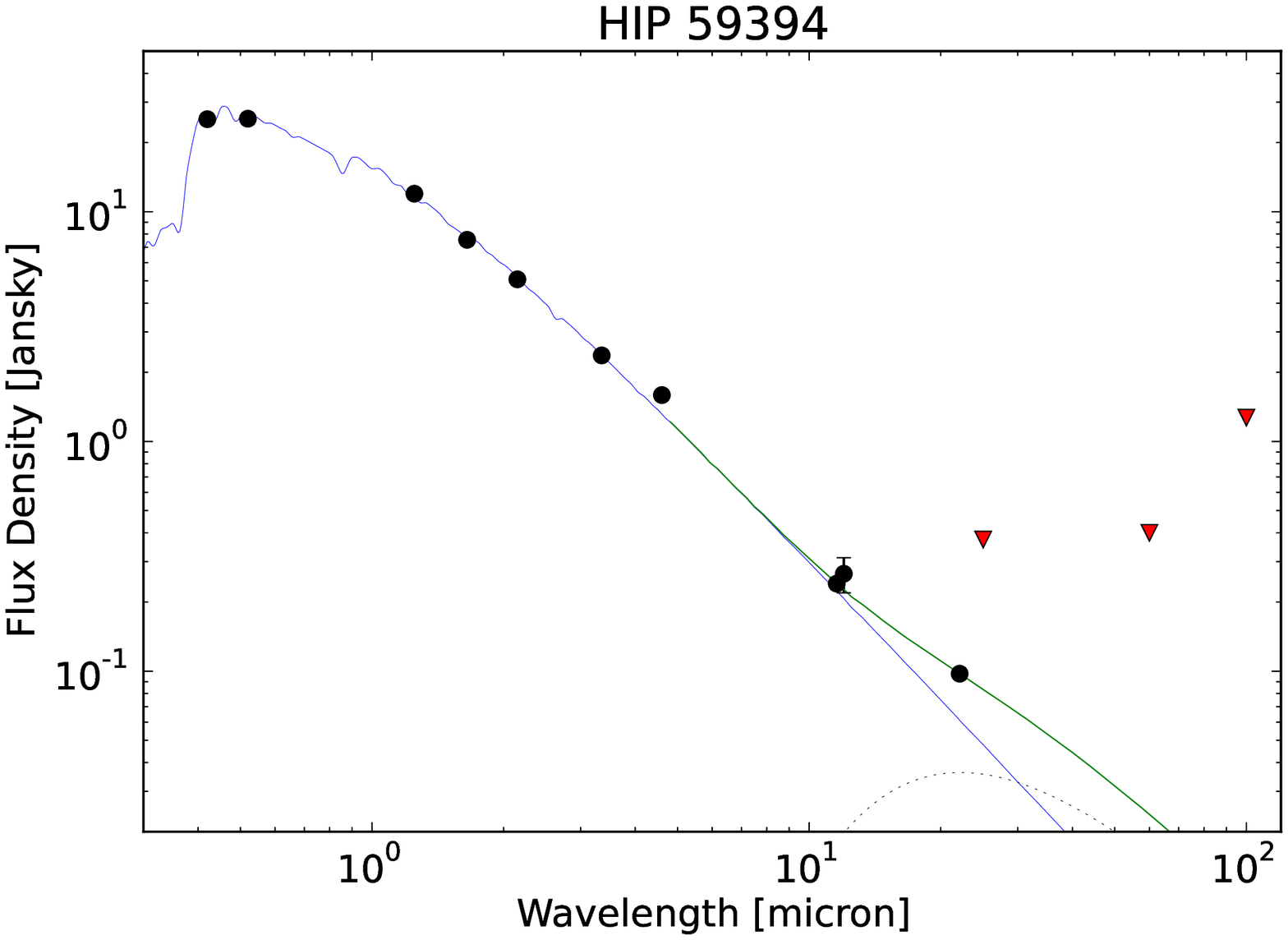} &
\includegraphics[scale=0.4]{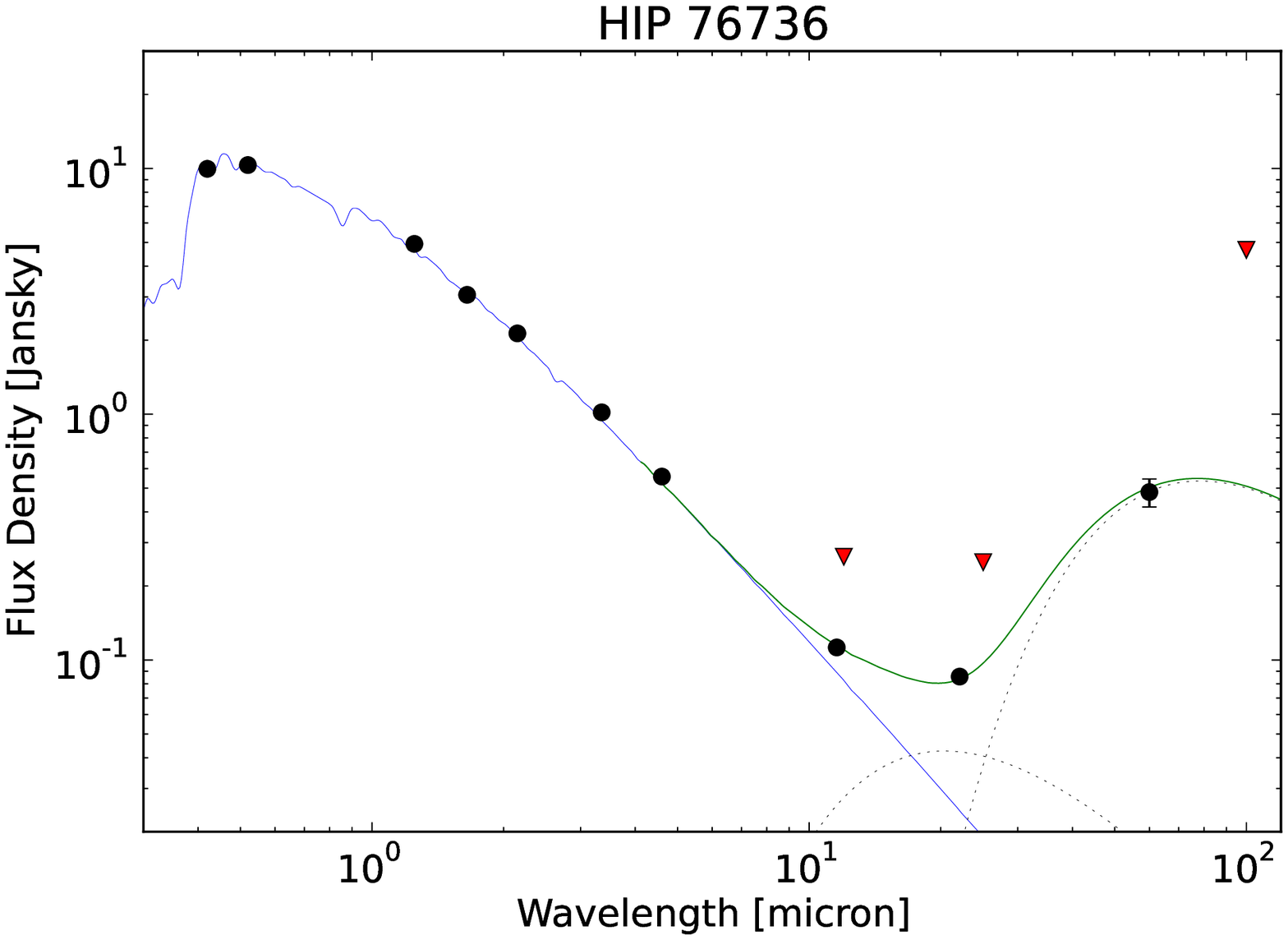} \\
\end{tabular}\par}
\caption{Spectral energy distribution of four stars in Tables 1 and 2 with excess emission at 22 $\mu$m wavelength in the WISE catalog; three of the four are considered in Section 3.  In each panel the two leftmost points are from Tycho-2, the next three points are
from 2MASS, and the next two are the two short wavelength channels in WISE.  The points at 11 and 12 $\mu$m are from WISE and IRAS, respectively.  Similarly the points at 22 and 25 $\mu$m are from WISE and IRAS.  Points at 60 and 100 $\mu$m are from IRAS.  IRAS upper limit flux densities are plotted as downward pointing triangles.  The IRAS points are color-corrected.  The blue line is a fit to the stellar photosphere (see Zuckerman et al 2011; Rhee et al 2007) and the green line is a sum of the photosphere and the excess emission from dust particles.  In each case the excess emission is modeled as a blackbody, or two blackbodies in the case of HIP 76736.  These blackbodies are drawn on the figures and have the following temperatures: 220 K, 240 K, 230 K, and 250 K and 65 K for HIP 26395, 41081, 59394 and 76736, respectively.  The total fractional infrared dust luminosities, $\tau$, are 8.8, 7.3, 4.4, and 59 x 10$^{-5}$ for HIP 26395, 41081, 59394, and 76736, respectively.}
\end{figure}

%\clearpage
%Spectral energy distributions (SEDs) for AB Dor stars in Table 8 with probable or definite infrared excess emission.  Near infrared JHK data points are from the 2MASS catalog.  Square data points between 3.5 and 8 $\mu$m are from the IRAC camera on Spitzer.  Diamonds at 9 and 18 $\mu$m are from AKARI.  Triangles at 24 and 70 $\mu$m are from the MIPS camera on Spitzer.  Circles between 12 and 100 $\mu$m are from IRAS.  The IRS spectrum from Spitzer is plotted at mid-IR wavelengths.

%\clearpage
%\begin{figure}
%\includegraphics[width=140mm]{Figure2.eps}
%\caption{\label{figure2} Spectral energy distributions (SEDs) for AB Dor stars in Table 8 with probable or definite infrared excess emission.}
%\end{figure}

\end{document}